\documentclass[aps,prc,showpacs,twocolumn,superscriptaddress]{revtex4}
\usepackage{multirow}
\usepackage{graphicx}
\usepackage{subfigure}
\usepackage{color}
\usepackage{amsmath}
\usepackage{amssymb}
\usepackage[colorlinks=true]{hyperref}
\usepackage{tikz}
\usepackage{etoolbox}
\usepackage{bm}
\usepackage[utf8]{inputenc}
\usepackage{cleveref}
\usepackage{amsmath}
\usepackage{amssymb}
\usepackage{float}
\makeatletter
\renewcommand{\@thesubfigure}{\hskip\subfiglabelskip}
\makeatother

\definecolor{dgreen}{cmyk}{1.,0.,1.,0.4}        % dark green
\definecolor{orange}{cmyk}{0.,0.353,1.,0.}    % orange

\begin{document}
\title{Investigations on mixed harmonic cumulants in heavy-ion collisions at the LHC}

\author{Ming Li}
\affiliation{Department of Physics and State Key Laboratory of Nuclear Physics and Technology, Peking University, Beijing 100871, China}
\affiliation{Collaborative Innovation Center of Quantum Matter, Beijing 100871, China}

\author{You Zhou}
\email{you.zhou@cern.ch}
\affiliation{Niels Bohr Institute, University of Copenhagen, Blegdamsvej 17, 2100 Copenhagen, Denmark}

\author{Wenbin~Zhao}
\affiliation{Department of Physics and State Key Laboratory of Nuclear Physics and Technology, Peking University, Beijing 100871, China}
\affiliation{Institute of Particle Physics and Key Laboratory of Quark and Lepton Physics (MOE), Central China Normal University, Wuhan, Hubei 430079, China}

\author{Baochi Fu}
\affiliation{Department of Physics and State Key Laboratory of Nuclear Physics and Technology, Peking University, Beijing 100871, China}
\affiliation{Collaborative Innovation Center of Quantum Matter, Beijing 100871, China}

\author{Yawen Mou}
\affiliation{Department of Physics and State Key Laboratory of Nuclear Physics and Technology, Peking University, Beijing 100871, China}
\affiliation{Collaborative Innovation Center of Quantum Matter, Beijing 100871, China}

\author{Huichao~Song}
\email{huichaosong@pku.edu.cn}
\affiliation{Department of Physics and State Key Laboratory of Nuclear Physics and Technology, Peking University, Beijing 100871, China}
\affiliation{Collaborative Innovation Center of Quantum Matter, Beijing 100871, China}
\affiliation{Center for High Energy Physics, Peking University, Beijing 100871, China}

\date{\today}

\begin{abstract}
A series of new flow observables mixed harmonic multi-particle cumulants ($MHC$), which allow for the first time to quantify the correlations strength between different order of flow coefficients with various moments, was investigated using hydrodynamic model. These new observables are constructed based on multi-particle cumulants, and thus by design will be less sensitive to the non-flow contaminations. In addition to the previous study of correlation involving two flow coefficients with their second moments, both correlations of three flow coefficients and the correlations of higher order moments of $v_2$ and $v_3$ are systematically investigated using {\tt iEBE-VISHNU} hybrid model with two different initial conditions, {\tt AMPT} and {\tt TRENTo}, respectively. These systematic studies using hydrodynamic models will significantly improve the understanding on the correlations between different orders (and moments) of flow coefficients. The hydrodynamic predictions shown in this paper and the future comparisons to experimental measurements will provide more constraints on theoretical models and extract more information about the transport properties of the quark-gluon plasma created in heavy-ion collisions.
\end{abstract}

%% keywords here, in the form: keyword \sep keyword
%% PACS codes here, in the form: \PACS code \sep code
%% MSC codes here, in the form: \MSC code \sep code
%% or \MSC[2008] code \sep code (2000 is the default)


\maketitle

%-==========================================================================
\clearpage

\section{Introduction}
\label{section1}

The ultra-relativistic collisions of heavy nuclei at Relativistic Heavy Ion Collider (RHIC) and the Large Hadron Collider(LHC) provide the experimental avenues to produce the Quark-Gluon Plasma (QGP) at extreme densities and temperatures \cite{Gyulassy:2004zy,Muller:2006ee,Muller:2012zq}. The observed strong collective flow plays a major role in probing the properties of the fluid-like QGP \cite{Ackermann:2000tr,Adler:2003kt,Adams:2005ph,Alver:2008aa,Aamodt:2010pa,Chatrchyan:2012ta,
Voloshin:2008dg,Snellings:2011sz,Heinz:2013th,Luzum:2012da,Song:2017wtw}. The harmonics flow, which is driven by the initial spatial anisotropy of the overlapping region between two colliding nuclei and reflects the anisotropic expansion of the emitted particles in momentum space, can be characterized by~\cite{Voloshin:1994mz}:
\begin{equation}
P(\varphi)=\frac{1}{2\pi}\sum_{n=-\infty}^\infty\overrightarrow{V_n}e^{-in\varphi},
\end{equation}
where $\overrightarrow{V_n}$ is the $n$-th order flow-vector defined as $\overrightarrow{V_n}=v_ne^{in\Psi_n}$. $v_n$ is the $n$-th order anisotropic flow harmonics, calculated by
\begin{equation}
v_n=\left\langle \cos\left[n\left(\varphi-\Psi_n\right)\right]\right\rangle,
\label{defination of vn}
\end{equation}
where $\Psi_n$ is the $n$-th order flow symmetry plane angle and $\left\langle ...\right\rangle$ denotes the average over all particles in an event.

%DIF < ~\cite{Borghini:2000sa,Borghini:2001vi,Bilandzic:2010jr}
As the flow symmetry plane $\Psi_n$ is unknown in experiments, $v_n$ coefficient cannot be measured directly~\cite{Luzum:2012da,Song:2017wtw}. One practical approach is the multi-particle azimuthal correlations method~\cite{Borghini:2000sa,Borghini:2001vi,Bilandzic:2010jr} that extracts $v_n$ in terms of its moments. The $v_n$ coefficients as well as its higher order cumulants of $v_n\{m\}$ have been measured systematically in experiment \cite{ALICE:2011ab,Adam:2016izf,Aaboud:2017blb,Acharya:2018zuq,Acharya:2018ihu,Aad:2019xmh,Aaboud:2018ves,ATLAS:2012at,Aad:2013xma,Aad:2015lwa,Aad:2015gqa,Aaboud:2016yar,Adare:2008ae,Aad:2012gla,Aad:2014lta,ABELEV:2013wsa,Adler:2002pu} and extensively calculated by hydrodynamics and hybrid model simulations \cite{Song:2010mg,Gale:2013da,Gale:2012rq,Song:2013gia,Ryu:2015vwa,Zhu:2015dfa,Denicol:2015nhu,Xu:2016hmp,McDonald:2016vlt,Zhao:2017yhj,Zhao:2017rgg,Ryu:2017qzn,Zhao:2019ehg,Zhao:2020pty,Everett:2020xug,Everett:2020yty}
in the last ten years. It was found that the flow coefficients are very sensitive to both initial state fluctuations and the transport coefficients of the expanding QGP.

To constrain the initial state models further and to achieve a better understanding of the hydrodynamic evolution of the QGP, several new flow observables have been measured in experiments and studied by model calculations, including the event-by-event $v_n$ distributions~\cite{Gale:2012rq,Aad:2013xma,Yan:2014afa,Zhou:2015eya}, the event-plane correlations~\cite{Aad:2014fla,Qiu:2012uy,Bhalerao:2013ina,Teaney:2013dta,Bhalerao:2014xra}, and the correlations between flow coefficients~\cite{Bhalerao:2014xra,Aad:2015lwa,ALICE:2016kpq,Niemi:2015qia,Giacalone:2016afq,Zhu:2016puf,Qian:2016pau}, the decorrelations of the flow vector~\cite{Heinz:2013bua,Gardim:2012im,Khachatryan:2015oea,Pang:2014pxa,Pang:2015zrq,Xiao:2015dma,Ma:2016fve,Acharya:2017ino} etc. The Symmetric Cumulant $SC(m,n)$, which quantifies the correlations between two flow coefficients with their second moments, $v_m^2$ and $v_n^2$, was proposed in Ref.~\cite{Bilandzic:2013kga} and measured by the ALICE Collaboration for the first time~\cite{ALICE:2016kpq}. It was found that $SC(m,n)$ is more sensitive to the medium properties than the individual flow harmonics \cite{ALICE:2016kpq,Acharya:2017gsw,Zhu:2016puf}. It was also realized that the normalized Symmetric Cumulants $NSC(m,n)$~\cite{ALICE:2016kpq,Acharya:2017gsw,Zhu:2016puf}, that evaluate the relative correlation strength between different flow harmonics, provide more direct information on the initial state correlations.

However, $SC(m,n)$ and $NSC(m,n)$ only focus on the correlations between two flow coefficients with their second moments. The study of general correlations among any flow coefficients or the correlations with higher-order moments could potentially provide additional information on the joint probability density function $(p.d.f.)$ of particle anisotropic expansion than existing observables. This can be done using mixed harmonic cumulant ($MHC$) developed via the generic recursive algorithm~\cite{Moravcova:2020wnf}. Here $MHC$ provides the possibilities to study the higher-order moment correlations between $v_m^k$ and $v_n^l$ as well as the correlations involving three different flow coefficients. By construction, $MHC$ does not contain symmetry plane correlations and is expected to be insensitive to non-flow contaminations.

In this paper, we perform systematic studies of mixed harmonic cumulants by utilizing the state-of-the-art {\tt iEBE-VISHNU} hybrid model with two different initial conditions, named {\tt AMPT} and {\tt TRENTo}. We will focus on exploring the sensitivities of the initial conditions and the QGP transport coefficients with $MHC$. The paper is organized as follows. Section \uppercase\expandafter{\romannumeral2} introduces the {\tt iEBE-VISHNU} hydrodynamic model with {\tt AMPT} and {\tt TRENTo} initial conditions and the setup of calculations. Section \uppercase\expandafter{\romannumeral3} presents our definition of $MHC$ and the methodology to calculate these new observables. Section \uppercase\expandafter{\romannumeral4} presents the results and discussions. Section \uppercase\expandafter{\romannumeral5} summarizes this paper.

%DIF < We will illustrate that $MHC$ has the potential to open a new window to explore the nature of the dynamic evolution of the created QGP in heavy-ion collisions. On the other hand, assuming that $v_{2}$ and $v_3$ are linearly correlated with the corresponding initial $\varepsilon_2$ and $\varepsilon_3$, then $MHC(\epsilon_2^k,\epsilon_3^p)$ from initial state is expected to be consistent with $MHC(v_2^k,v_3^p)$ from final state. This expectation will be examined using {\tt iEBE-VISHNU} hybrid model with two initial state models and different $\eta/s$ parameters.

\section{The model and set-ups}

In this paper, we implement {\tt iEBE-VISHNU} to study the novel mixed harmonic multi-particle cumulants in Pb+Pb collisions at $\sqrt{s_{NN}}$ = 5.02 TeV. {\tt iEBE-VISHNU}~\cite{Shen:2014vra} is an event-by-event hybrid model that combines (2+1)-d viscous hydrodynamics {\tt VISH2+1}~\cite{Song:2007fn,Song:2007ux,Song:2009gc} to describe the QGP expansion and the hadron cascades model, Ultra-relativistic Quantum Molecular Dynamics ({\tt UrQMD})~\cite{Bass:1998ca,Bleicher:1999xi} to simulate the evolution of hadronic matter. With a Bjorken approximation $v_z=z/t$~\cite{Bjorken:1982qr}, {\tt VISH2+1}~\cite{Song:2007fn,Song:2007ux,Song:2009gc}  solves the 2+1-d  transport equations for energy-momentum tensor $T^{\mu \nu}$ and the 2nd order Israel-Stewart equations for shear stress tensor $\pi^{\mu \nu}$ and bulk pressure $\Pi$ with a state-of-the-art equation of state (EoS) s95-PCE as an input~\cite{Huovinen:2009yb,Shen:2010uy}. At the switching temperature $T_{sw}$ closed to the phase transition, the hydrodynamics description of the QGP expansion is changed to the  hadron cascade simulations with various hadrons resonance generated from the freeze-out hyper-surface using the {\tt iSS} event generator based on the Cooper-Fryer formula~\cite{Song:2010aq}. Then  the subsequent evolution of  hadronic matter are simulated by the {\tt UrQMD} model~\cite{Bass:1998ca,Bleicher:1999xi}, which propagates produced hadrons along classical trajectories, together with the elastic scatterings, inelastic scatterings and resonance decays until all hadrons in the system cease interactions.

In the following calculations, we implement two different initial conditions, called {\tt AMPT} and {\tt TRENTo}, for the {\tt iEBE-VISHNU} simulations. For {\tt AMPT} initial condition, the initial energy density profiles are constructed by the energy decompositions of individual partons, where the fluctuation scales related to the energy decompositions can be changed by a Gaussian smearing factor $\sigma$ \cite{Bhalerao:2015iya,Pang:2012he,Xu:2016hmp,Fu:2020oxj}. Following~\cite{Xu:2016hmp}, we truncate the produced partons from  {\tt AMPT} within $|\eta|<1$ to construct the initial energy density profiles in the transverse plane at mid-rapidity. {\tt TRENTo} is a parameterized initial condition model based on the eikonal entropy deposition via a reduced thickness function~\cite{Moreland:2014oya}. By adjusting the entropy deposition parameter in the initial model, {\tt TRENTo} can interpolate among different initial condition models such as {\tt MC-Glauber}, {\tt MC-KLN}, {\tt EKRT} and so on \cite{Bernhard:2016tnd,Moreland:2014oya}. For more details about these two initial conditions, please refer to~\cite{Moreland:2014oya,Bernhard:2016tnd,Xu:2016hmp}.

For the simulations of {\tt iEBE-VISHNU} with {\tt AMPT} initial condition, we take two different values of the specific shear viscosity as a contrast, $\eta/s=0.08$ and $\eta/s=0.2$ and neglect the bulk viscosity. For {\tt iEBE-VISHNU} with {\tt TRENTo} initial condition model, we take $\eta/s(T)$ and $\zeta/s(T)$. Other parameters exactly coincide with our previous paper~\cite{Zhao:2017yhj}, which has nicely described various flow observables in 2.76 A TeV and 5.02 A TeV  Pb+Pb collisions.

\begin{figure*}[t]
\vspace{-0.3cm}
\centering
\includegraphics[width=1.0\linewidth]{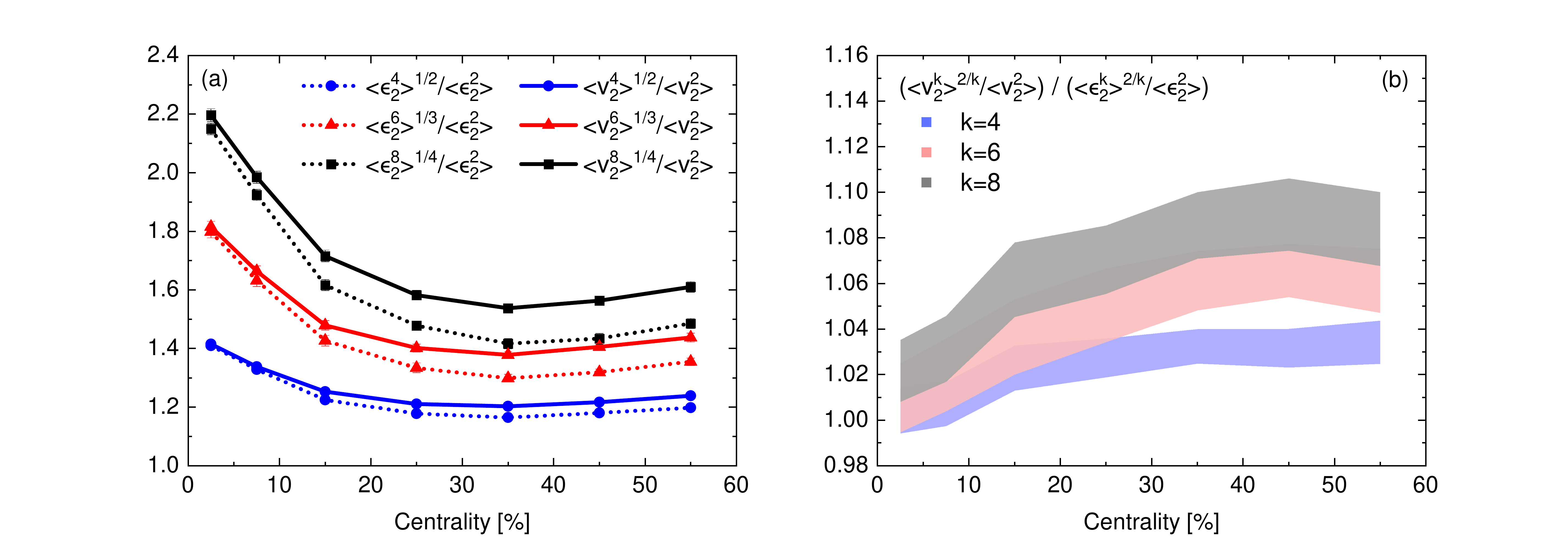}
\setlength{\abovecaptionskip}{-0.2cm}
\caption{(Color online) Centrality dependence of (a) $\left<v_{2}^{k}\right>^{2/k}/\left<v_{2}^{2}\right>$ compared with $\left<\epsilon_{2}^{k}\right>^{2/k}/\left<\epsilon_{2}^{2}\right>$, and (b) the ratios of $(\left<v_{2}^{k}\right>^{2/k}/\left<v_{2}^{2}\right>) / (\left<\epsilon_{2}^{k}\right>^{2/k}/\left<\epsilon_{2}^{2}\right>)$ for $k=4,6,8$, calculated by {\tt iEBE-VISH2+1} with {\tt AMPT} initial conditions and $\eta/s=0.08$.}
\label{fig1}
\end{figure*}

\section{CORRELATIONS BETWEEN DIFFERENT MOMENTS OF FLOW COEFFICIENT}
\label{sec:definition}
%DIF < which means the azimuthal angle is the fundamental observable of the cumulant.

As introduced in the introduction, the correlations between different order moments of various flow coefficients can be studied using multi-particle cumulants of mixed harmonics, called mixed harmonic cumulants, $MHC$~\cite{Moravcova:2020wnf}. It is constructed as the multi-particle cumulant in terms of the azimuthal angle of each emitted particle, following the traditional approach of constructing multi-particle cumulants~\cite{Borghini:2001vi, Bilandzic:2010jr}.

For the $MHC$ with 4-particles, it is equivalent to symmetric cummulant $SC(m,n)$, which is $MHC(v_n^2, v_m^2) = SC(m,n)$.  The 6-particle cumulants study the genuine correlation between two flow coefficients $(v_2^4, v_3^2)$, $(v_2^2, v_3^4)$ and correlations between three flow coefficients $(v_2^2, v_3^2, v_4^2)$ and $(v_3^2, v_4^2, v_5^2)$, which are defined as~\cite{Moravcova:2020wnf}:
\begin{eqnarray}
\begin{aligned}
MHC(v_2^4,v_3^2) =& \left\langle v_2^4 v_3^2\right\rangle - 4\left\langle v_2^2 v_3^2\right\rangle \left\langle v_2^2\right\rangle \nonumber \\
& -\left\langle v_2^4\right\rangle \left\langle v_3^2\right\rangle + 4\left\langle v_2^2\right\rangle^2 \left\langle v_3^2\right\rangle,  \\
\end{aligned}
%\label{eq:6pc}
\end{eqnarray}
\begin{eqnarray}
\begin{aligned}
MHC(v_2^2,v_3^4) =& \left\langle v_2^2 v_3^4\right\rangle - 4\left\langle v_2^2 v_3^2\right\rangle \left\langle v_3^2\right\rangle -\left\langle v_2^2\right\rangle \left\langle v_3^4\right\rangle \nonumber \\
&+ 4\left\langle v_2^2\right\rangle \left\langle v_3^2\right\rangle^2,  \\
\end{aligned}
\end{eqnarray}
\begin{eqnarray}
\begin{aligned}
MHC(v_2^2,v_3^2,v_4^2) =& \left\langle v_2^2 v_3^2 v_4^2\right\rangle - \left\langle v_2^2 v_3^2\right\rangle \left\langle v_4^2\right\rangle - \left\langle v_2^2 v_4^2\right\rangle \left\langle v_3^2\right\rangle \nonumber \\
&-\left\langle v_3^2 v_4^2\right\rangle \left\langle v_2^2\right\rangle + 2\left\langle v_2^2\right\rangle \left\langle v_3^2\right\rangle\left\langle v_4^2\right\rangle, \\
\end{aligned}
\end{eqnarray}
\begin{eqnarray}
\begin{aligned}
MHC(v_3^2,v_4^2,v_5^2) =& \left\langle v_3^2 \, v_4^2 \, v_5^2\right\rangle - \left\langle v_3^2 \, v_4^2\right\rangle \left\langle v_5^2\right\rangle - \left\langle v_3^2 v_5^2\right\rangle \left\langle v_4^2\right\rangle \nonumber \\
&-\left\langle v_4^2  v_5^2\right\rangle \, \left\langle v_3^2\right\rangle + 2\left\langle v_3^2\right\rangle \left\langle v_4^2\right\rangle \left\langle v_5^2\right\rangle.
\end{aligned}
\end{eqnarray}
%Here all the lower order correlations (second and fourth orders) have been subtracted from the 6-particle correlation.

For MHC with 8-particles, the three different MHCs to study the correlations between $(v_2^6, v_3^2)$, $(v_2^4, v_3^4)$ and $(v_2^2, v_3^6)$ are defined as:
\begin{eqnarray}
\begin{aligned}
MHC(v_2^6,v_3^2) =& \left\langle v_2^6 v_3^2\right\rangle -9\left\langle v_2^4 v_3^2\right\rangle \left\langle v_2^2\right\rangle -\left\langle v_2^6\right\rangle \left\langle v_3^2\right\rangle \nonumber \\
&-9\left\langle v_2^4\right\rangle \left\langle v_2^2 v_3^2\right\rangle -36\left\langle v_2^2\right\rangle^3\left\langle v_3^2\right\rangle  \nonumber \\
&+18\left\langle v_2^2\right\rangle \left\langle v_3^2\right\rangle \left\langle v_2^4\right\rangle +36\left\langle v_2^2\right\rangle^2\left\langle v_2^2 v_3^2\right\rangle \\
\end{aligned}
%\label{eq:8pc}
\end{eqnarray}
\begin{eqnarray}
\begin{aligned}
MHC(v_2^4,v_3^4)=& \left\langle v_2^4 v_3^4\right\rangle -4\left\langle v_2^4 v_3^2\right\rangle \left\langle v_3^2\right\rangle \nonumber \\
&-4\left\langle v_2^2 v_3^4\right\rangle \left\langle v_2^2\right\rangle -\left\langle v_2^4\right\rangle \left\langle v_3^4\right\rangle \nonumber \\
&-8\left\langle v_2^2 v_3^2\right\rangle^2 -24\left\langle v_2^2\right\rangle^2 \left\langle v_3^2\right\rangle^2 \nonumber \\
&+4\left\langle v_2^2\right\rangle^2 \left\langle v_3^4\right\rangle +4\left\langle v_2^4\right\rangle \left\langle v_3^2\right\rangle^2 \nonumber \\
&+32\left\langle v_2^2\right\rangle \left\langle v_3^2\right\rangle \left\langle v_2^2 v_3^2\right\rangle \\
\end{aligned}
\end{eqnarray}
\begin{eqnarray}
\begin{aligned}
MHC(v_2^2,v_3^6)=& \left\langle v_2^2 v_3^6\right\rangle -9\left\langle v_2^2 v_3^4\right\rangle \left\langle v_3^2\right\rangle -\left\langle v_3^6\right\rangle \left\langle v_2^2\right\rangle \nonumber \\
&-9\left\langle v_3^4\right\rangle \left\langle v_2^2 v_3^2\right\rangle -36\left\langle v_2^2\right\rangle \left\langle v_3^2\right\rangle^3  \nonumber \\
&+18\left\langle v_2^2\right\rangle \left\langle v_3^2\right\rangle \left\langle v_3^4\right\rangle +36\left\langle v_3^2\right\rangle^2 \left\langle v_2^2 v_3^2\right\rangle \\
\end{aligned}
\end{eqnarray}
These $MHC$ observables are constructed via multi-particle correlations, which can be easily calculated via the newly proposed generic algorithm~\cite{Moravcova:2020wnf}, a more advanced approach with exact and efficient calculations. For more details, please refer to Ref.~\cite{Moravcova:2020wnf}. Considering the limited statistics of hydrodynamic simulations, we do not extend the list to higher-order $MHC$ in this paper.

To evaluate the relative strength of the correlations between different flow harmonics and remove the dependence on the magnitude of flow coefficients, one defines the normalized mixed harmonic cumulants $nMHC$ by normalizing the $MHC$ with flow coefficients. These normalized mixed harmonic cumulants  provide a  quantitative comparison between different collision energies, and between the initial state and final state correlations in model simulations. The $nMHC$ with two and three flow coefficients are defined as:
\begin{eqnarray}
&& nMHC(v_m^k,v_n^l)=\frac{MHC(v_m^k,v_n^l)}{\left\langle v_m^k\right\rangle\left\langle v_n^l\right\rangle},
\label{def_nMHC} \\
&& nMHC(v_m^k,v_n^l, v_p^q)=\frac{MHC(v_m^k,v_n^l, v_p^q)}{\left\langle v_m^k\right\rangle \left\langle v_n^l\right\rangle\left\langle v_p^q\right\rangle}
\label{nMHCH_3h}
\end{eqnarray}

\begin{figure*}[t] %Fig.2
\vspace{-0.8cm}
\centering
\includegraphics[width=1.0\textwidth]{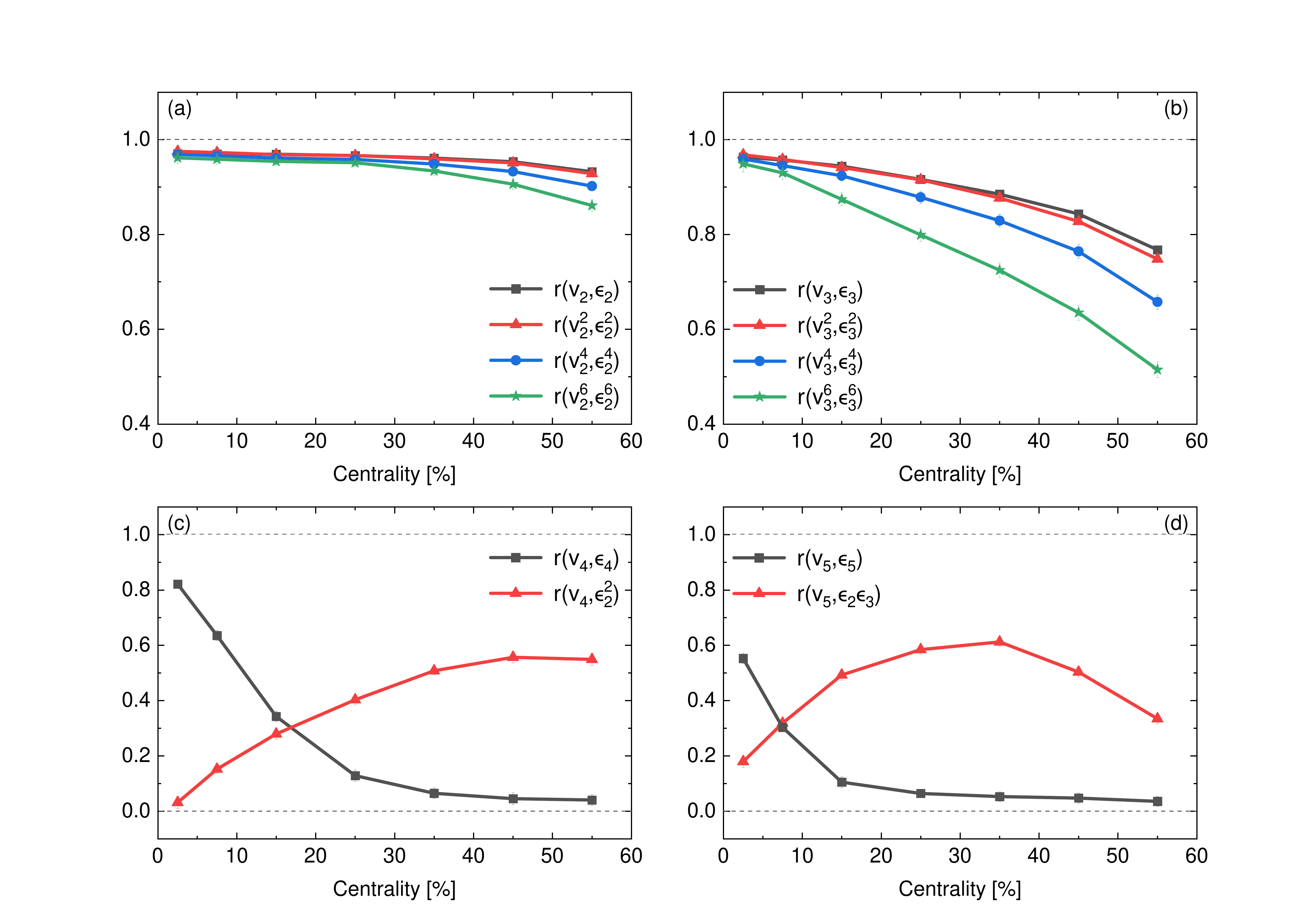}
\setlength{\abovecaptionskip}{-0.3cm}
\caption{(Color online) The Pearson correlation coefficients between flow harmonics and initial eccentricity in Pb+Pb collisions at $\sqrt{s_{NN}}$=5.02 TeV calculated by {\tt iEBE-VISH2+1} with {\tt AMPT} initial conditions and $\eta/s=0.08$.}
\label{fig:vnen}
\end{figure*}
%with less CPU demanding. For the single event $m-$particle correlations, it can be calculated by calling function {\tt complex Correlator} developed in the generic algorithm,

%Assuming that $v_n \propto \varepsilon_n$, then one
We also study the correlations between two and three different initial anisotropy coefficients in various moments in an analogous way:
\begin{eqnarray}
&& nMHC(\epsilon_m^k, \epsilon_n^l)=\frac{MHC(\epsilon_m^k, \epsilon_n^l)}{\left\langle \epsilon_m^k\right\rangle\left\langle \epsilon_n^l\right\rangle} \\
&& nMHC(\epsilon_m^k,\epsilon_n^l, \epsilon_p^q)=\frac{MHC(\epsilon_m^k,\epsilon_n^l, \epsilon_p^q)}{\left\langle \epsilon_m^k\right\rangle \left\langle \epsilon_n^l\right\rangle\left\langle \epsilon_p^q\right\rangle}
\label{eq:MHC_en}
\end{eqnarray}
where $\epsilon_n$ is the $n$-th eccentricity defined as:
\begin{equation}
\epsilon_n e^{in\Phi_n}=-\frac{\int rdrd\varphi r^n e^{in\varphi}\varepsilon(r,\varphi)}{\int rdrd\varphi r^n \varepsilon(r,\varphi)}.
\label{defination of eccentricity}
\end{equation}
Here $\Phi_n$ is the initial symmetry plane and $\varepsilon(r,\varphi)$ is the initial energy density in the transverse plane~\cite{Qiu:2011iv,Shen:2014vra}.

\begin{figure*}[t] %Fig.3
\vspace{-1.5cm}
\centering
\includegraphics[width=1.0\linewidth]{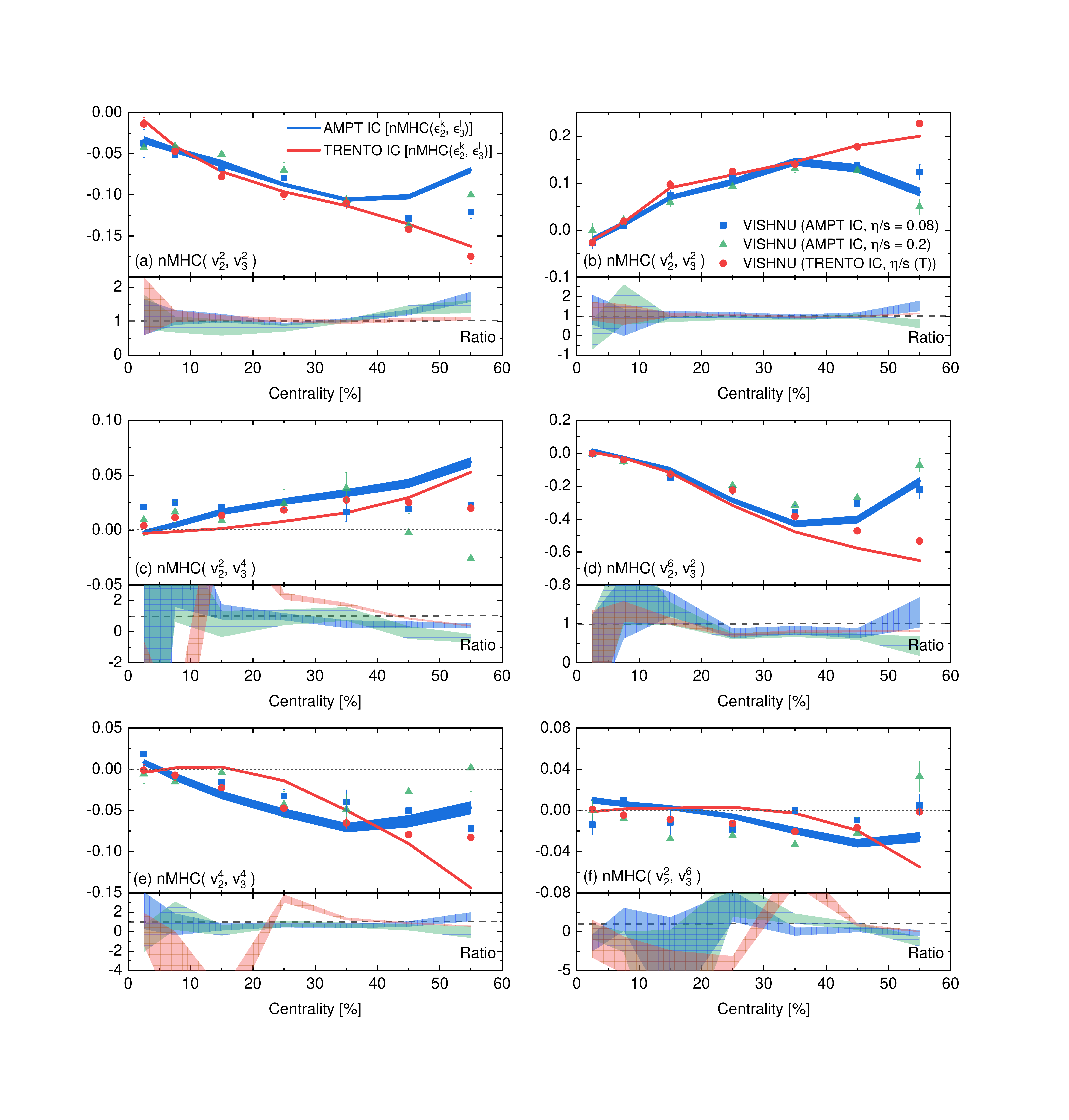}
\setlength{\abovecaptionskip}{-1.7cm}
\caption{(Color online) The $nMHC(v_2^k,v_3^l)$ are calculated by {\tt iEBE-VISHNU} with {\tt AMPT} and {\tt TRENTo} initial conditions in Pb+Pb collisions at $\sqrt{s_{NN}}$=5.02 TeV. The $nMHC(\epsilon_2^k,\epsilon_3^l)$ are calculated by {\tt AMPT} and {\tt TRENTo} initial conditions respectively. The Ratios are defined by $nMHC(v_2^k,v_3^l)/nMHC(\epsilon_2^k,\epsilon_3^l)$.}
\label{fig:nMHC23}
\end{figure*}

\section{Results and Discussion}
\label{sec:result}

\begin{figure*}[t] %Fig 4
\vspace{-1.2cm}
\centering
\includegraphics[width=1.0\textwidth]{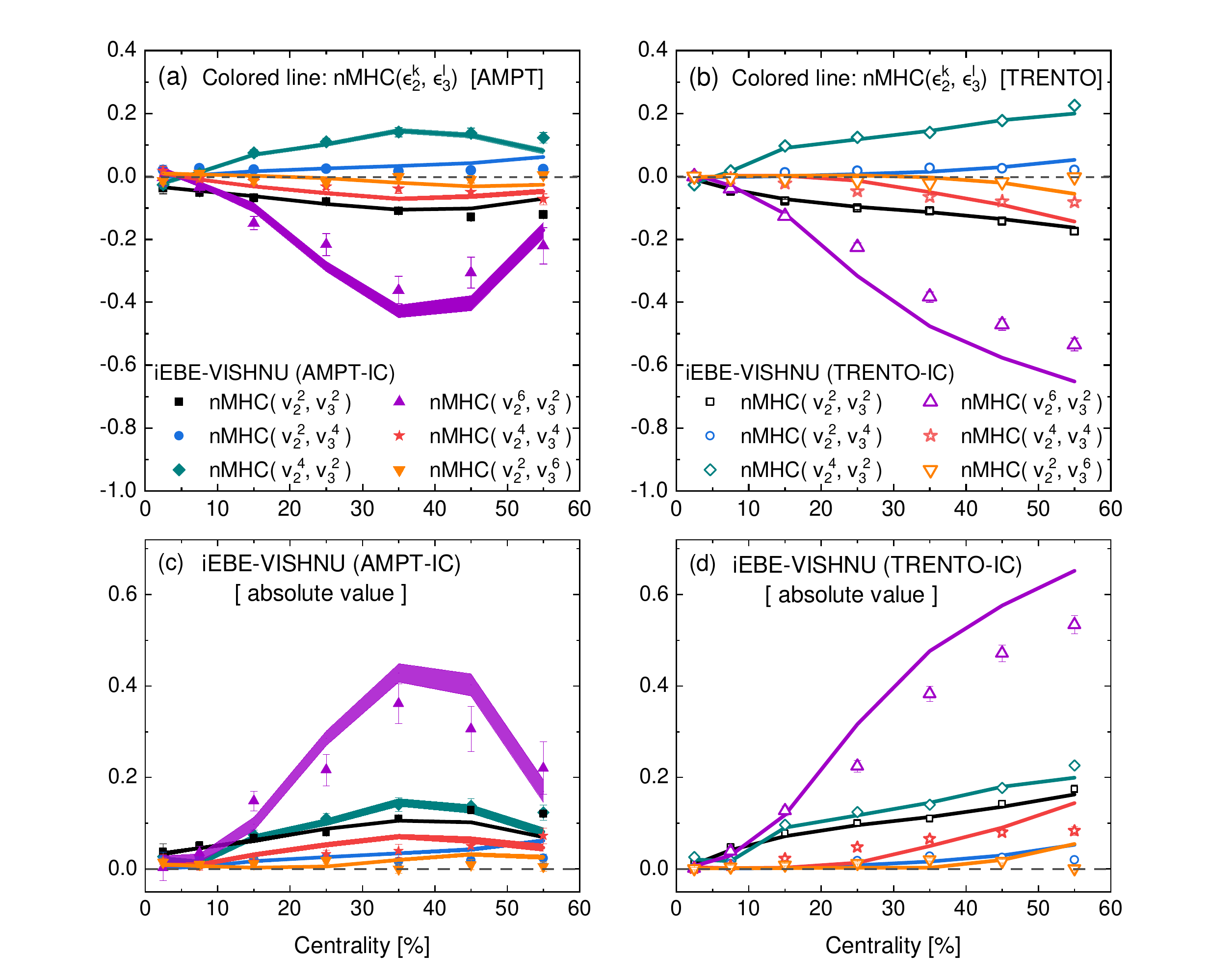}
\setlength{\abovecaptionskip}{-0.3cm}
\caption{(Color online) The $nMHC(v_2^k,v_3^l)$ and $nMHC(\epsilon_2^k,\epsilon_3^l)$ from  {\tt AMPT} initial model (a)(c) and {\tt TRENTo} initial model (b)(d) in Pb+Pb collisions at $\sqrt{s_{NN}}$=5.02 TeV. The lower panels show absolute values of $nMHC(v_2^k,v_3^l)$ and $nMHC(\epsilon_2^k,\epsilon_3^l)$.}
\label{fig:ordering}
\end{figure*}

\begin{figure*}[t]
\vspace{-0.95cm}
\centering
\includegraphics[width=1.0\textwidth]{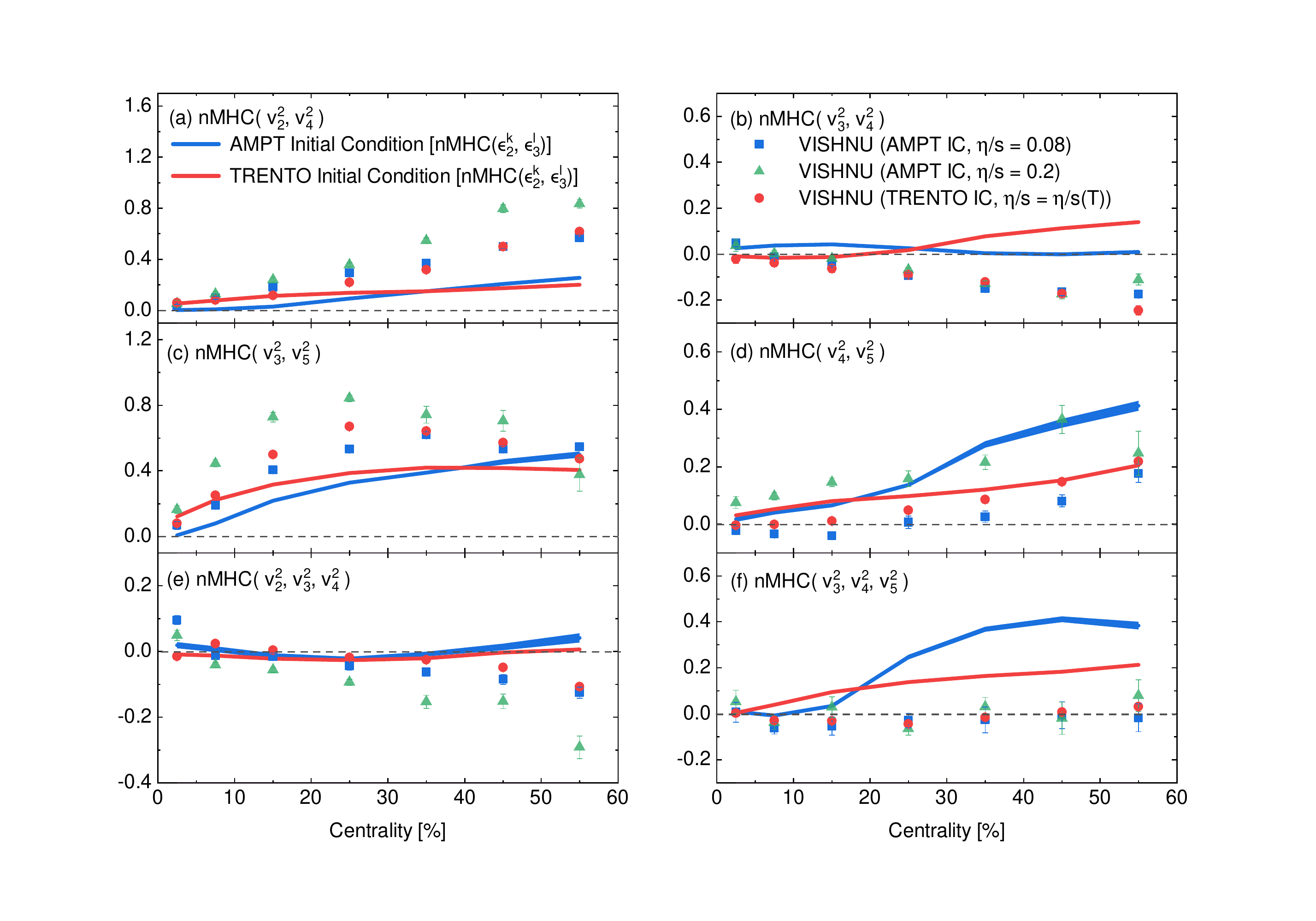}
\setlength{\abovecaptionskip}{-1.1cm}
\caption{(Color online) The $nMHC(v_m^k,v_n^l)$ and $nMHC(v_m^k,v_n^l,v_p^q)$ are calculated by {\tt iEBE-VISHNU} with {\tt AMPT} and {\tt TRENTo} initial conditions respectively in Pb+Pb collisions at $\sqrt{s_{NN}}$=5.02 TeV. The $nMHC(\epsilon_m^k,\epsilon_n^l)$ and  $nMHC(\epsilon_m^k,\epsilon_n^l,\epsilon_p^q)$ are calculated by {\tt AMPT} and {\tt TRENTo} initial conditions respectively.}
\label{fig:nMHCNL}
\end{figure*}

As discussed earlier, the correlations between flow coefficients have a unique sensitivity to the initial conditions as well as the properties of the QGP created in relativistic heavy-ion collisions. Based on the assumption that $v_n \propto \epsilon_n$, one can use normalized symmetric cumulants in the final state to directly constrain the initial conditions by the normalized symmetry cumulant in term of anisotropy coefficients. However, it has been presented by a series of works~\cite{Gardim:2011xv,Teaney:2010vd,Noronha-Hostler:2015dbi,Niemi:2012aj,Gardim:2014tya,Plumari:2015cfa,Fu:2015wba} that such linear response of final state $v_n$ to the corresponding initial $\epsilon_n$ does not hold in general. This non-linear response plays a non-negligible role in the correlations between flow coefficients. For lower order flow coefficients, it was found that $v_2$ and $v_3$ have approximately linear response to the corresponding initial eccentricities in central and middle central collisions, based on the study of correlation between $v_{2,3}$ and $\epsilon_{2,3}$ with $c(v_{2,3}, \epsilon_{2,3})$~\cite{Niemi:2012aj}, as well as using the comparisons between $v_{2,3}\{4\}/v_{2,3}\{2\}$ and $\epsilon_{2,3}\{4\}/\epsilon_{2}\{2\}$~\cite{Giacalone:2017uqx}. For higher-order anisotropic flow such as $v_4$ and $v_5$, the nonlinear hydrodynamic response can not be neglected \cite{Teaney:2010vd,Niemi:2012aj,Bhalerao:2014xra,Qian:2016fpi,Yan:2015jma} and have been investigated in great details in experiments.

Besides what have been done so far, the effects of non-trivial non-linear response in $v_2$ and $v_{3}$ might be enhanced in the correlations of $v_{2,3}$ and $\epsilon_{2,3}$ with their higher order moments. The result presented in Fig.~\ref{fig1} indeed shows a larger deviation between $\left< v_{2}^{k} \right>^{2/k}/\left< v_{2}^{2} \right>$ and $\left< \epsilon_{2}^{k} \right>^{2/k}/\left< \epsilon_{2}^{2} \right>$ for $k=8$ than $k=4$. It suggests that higher-order moments of $v_{n}$ are more sensitive to the non-linear response. On the other hand, the possible study of multi-particle cumulants $v_{2,3}\{k\}/v_{2,3}\{2\}$ for $k=6, 8$ might not bring additional information to $v_{2,3}\{4\}/v_{2,3}\{2\}$ because the expected probability density function of $v_n$ coefficient gives $v_{2,3}\{4\} \approx v_{2,3}\{6\} \approx v_{2,3}\{8\}$~\cite{Aad:2014vba,Sirunyan:2017fts,Acharya:2018lmh}. %[Maybe useful to highlight] However, this study is not practical in experiments due to non-flow contaminations in multi-particle correlations measurements.

Alternative way to quantitatively evaluate the linear response of the final state $v_n$ to the initial state $\epsilon_n$ is using the Pearson correlation coefficient~
\cite{Gardim:2014tya,Niemi:2012aj,Plumari:2015cfa,Noronha-Hostler:2015dbi}:
\begin{equation}
r(X,Y)=\frac{\left\langle XY\right\rangle-\left\langle X\right\rangle\left\langle Y\right\rangle}{\sqrt{Var[X]}\sqrt{Var[Y]}}
\end{equation}
where $Var[X]$ is the variance of X. Generally speaking, the closer the value of $r(X,Y)$ approaching 1 stands for the  higher the degree of linear correlation between $X$ and $Y$. By inserting $v_n^{k}$ and $\epsilon_n^{k}$ into the above equations, one can study the linear correlation of $v_n^k$ and $\epsilon_{n}^{k}$ with $r(v_n^{k}, \epsilon_n^{k})$.

In this paper, the $r(v_n^{k}, \epsilon_n^{k})$ calculations are calculated based on simulations of {\tt iEBE-VISH2+1} model with {\tt AMPT} initial conditions, where the specific shear viscosity takes $\eta/s=0.08$. The results are presented in Fig.~\ref{fig:vnen}. It can be seen in Fig.~\ref{fig:vnen} (a) that $r(v_2,\epsilon_2)$ are close to 1 in central and semi-central collisions, which suggests linear correlations between $v_2$ and $\epsilon_2$. For the $r(v_3,\epsilon_3)$ shown in Fig.~\ref{fig:vnen} (b), the linear correlation of $v_3$ and $\epsilon_3$ gradually breaks down from central to peripheral collisions. Apparently, the linear response of $v_3$ to $\epsilon_3$ is worse than $v_{2}$ to $\epsilon_2$. These results confirm the previous observations based on different hydrodynamic models~\cite{Gardim:2014tya,Niemi:2012aj,Plumari:2015cfa,Noronha-Hostler:2015dbi}. Beside the study of correlations between $v_2$ ($v_3$) and $\epsilon_2$ ($\epsilon_3$), it is demonstrated in Fig.~\ref{fig:vnen} (a) and (b) that $r(v_3^k,\epsilon_3^k)$ shows stronger deviation from unity than $r(v_2^k,\epsilon_2^k)$ for other orders ($k = 2, 4, 6$). Such results imply that for no matter which order of $k$, the linear response of $v_2^{k}$ to $\epsilon_2^k$ holds better than $v_3^k$ to $\epsilon_3^k$ in non-central collisions.
%Instead of comparing $r(v_2^k,\epsilon_2^k)$ and $r(v_3^k,\epsilon_3^k)$ in each order of $k$, Fig.~\ref{fig:vnen} (e) and (f) group the results of $r(v_2^k,\epsilon_2^k)$ and $r(v_3^l,\epsilon_3^l)$ with different orders, respectively.
It can be also found that $r(v_2^k,\epsilon_2^k)$ with different orders are close to the unity, with a very weak centrality dependence. In contrast, the deviations of $r(v_3^l,\epsilon_3^l)$ from unity are getting stronger from central to peripheral collisions, such deviation are more pronounced for higher order $r(v_3^l,\epsilon_3^l)$.
Besides, Fig.~\ref{fig:vnen} (c) shows that $v_4$ is almost linearly correlated with $\epsilon_4$ in central collisions, such correlations decreases rapidly with the centrality increasing. On the other hand, the correlations between $\epsilon_2^2$ to $v_4$ become more important from semi-central collisions. A similar study is performed for $v_5$ in Fig.~\ref{fig:vnen} (d). It shows that the linear relationship between $\epsilon_5$ and $v_5$ holds better in very central collisions, while the correlation between $\epsilon_2\epsilon_3$ and $v_5$ shows a non-monotonic centrality dependence. Such observations are mostly explained by the nonlinear hydrodynamic response of higher-order anisotropy flow to the initial conditions.

The study of correlations between initial $\epsilon_n$ and final state $v_n$ have been discussed in great detail in Fig.~\ref{fig:vnen}. It undoubtedly shows that the linear relationship between them carry important information on the initial conditions using $v_2$ for the presented centrality ranges or using $v_3$ in the central collision, while nonlinear correlations shed new light into the hydrodynamic evolution of the created matter, e.g., using $v_3$ in non-central collisions. Despite the importance of the above-mentioned study with Pearson coefficients between $v_n^k$ and $\epsilon_n^k$, it can not be done experimentally. Instead, one can apply the newly proposed $nMHC$ to study the linear and nonlinear responses. %In this paper, we firstly investigate the different order of $nMHC(v_2^k,v_3^l)$ within the framework of {\tt iEBE-VISHNU} with {\tt AMPT} and {\tt TRENTo} initial conditions in Pb+Pb collisions at $\sqrt{s_{NN}}$ = 5.02 TeV.

To study the sensitivity of the $nMHC$ to the $\eta/s$ values, two different specific shear viscosity with $\eta/s=0.08$ and 0.2 are used for the {\tt iEBE-VISHNU} simulations with {\tt AMPT} initial conditions. The results of $nMHC(v_2^k,v_3^l)$ of the 4-particle, 6-particle and 8-particle cumulants are shown in Fig.~\ref{fig:nMHC23}. Consistent results of $nMHC(v_2^2,v_3^2)$ from two different $\eta/s$ values have been seen in the centrality region of 0-50\%, which suggests that $nMHC(v_2^2,v_3^2)$ is insensitive to $\eta/s$. In addition, there is a sizeable difference between the calculations from {\tt AMPT} and {\tt TRENTo} initial conditions, which shows the sensitivity of $nMHC(v_2^2,v_3^2)$ to the initial correlations between $\epsilon_2^2$ and $\epsilon_3^2$. This observation is not a surprise as $nMHC(v_2^2,v_3^2)$ is practically identical to $NSC(3,2)$ which have been carefully studied before~\cite{Zhu:2016puf}.
The compatible values of $nMHC(v_2^4,v_3^2)$ calculated by different $\eta/s$ from {\tt iEBE-VISHNU} with {\tt AMPT} initial conditions in the centrality region of 0-50\% shown in Fig.~\ref{fig:nMHC23} (b) suggest $nMHC(v_2^4,v_3^2)$ is insensitive to $\eta/s$. Besides, $nMHC(v_2^4,v_3^2)$ shows no sensitivity to initial conditions in central and semi-central collisions but presents the difference in peripheral collisions. Meanwhile, $nMHC(\epsilon_2^4,\epsilon_3^2)$ also shows sensitivity to initial conditions in peripheral collisions. All the models show the ratios $nMHC(v_2^4,v_3^2)$/$nMHC(\epsilon_2^4,\epsilon_3^2)$ are close to 1 within the allowed error range, which demonstrates the linear response from final state to initial state is tenable in this case.
Fig.~\ref{fig:nMHC23} (c) shows that for the results from {\tt iEBE-VISHNU} with {\tt AMPT} initial conditions, $nMHC(v_2^2,v_3^4)$ shows no sensitivity to $\eta/s$ in non-peripheral collisions. Besides, the results calculated by {\tt iEBE-VISHNU} with {\tt TRENTo} initial conditions are compatible with that from {\tt iEBE-VISHNU} with {\tt AMPT} initial conditions in non-peripheral collisions.
In addition, different initial conditions give very different values of $nMHC(\epsilon_2^2,\epsilon_3^4)$ in non-central collisions. The ratios deviate from 1, which suggests the linear response basically breaks down in this case.
The consistent results of $nMHC(v_2^6,v_3^2)$ from two different $\eta/s$ values in the centrality region of 0-50\% are shown in Fig.~\ref{fig:nMHC23} (d), which suggests $nMHC(v_2^6,v_3^2)$ shows no sensitivity to $\eta/s$. Besides, the results calculated by {\tt iEBE-VISHNU} with different initial conditions give consistent values in centra and semi-centra collisions but present a big difference in peripheral collisions. The $nMHC(\epsilon_2^6,\epsilon_3^2)$ calculated by different initial conditions are basically compatible in centra and semi-central collisions but show a sizable difference in peripheral collisions. The ratios are roughly close to 1 in central collisions, which suggests the linear response holds only for central collisions.
Fig.~\ref{fig:nMHC23} (e) shows $nMHC(v_2^4,v_3^4)$ is insensitive to $\eta/s$ in centra and semi-centra collisions but sensitive to initial conditions from semi-central to peripheral collisions. Different initial condition models show a sizable difference of $nMHC(\epsilon_2^4,\epsilon_3^4)$. The ratios totally deviate from 1 especially for the calculations from {\tt iEBE-VISHNU} with {\tt TRENTo} initial conditions, which suggests the linear response totally breaks down in this case.
Fig.~\ref{fig:nMHC23} (f) shows $nMHC(v_2^2,v_3^6)$ is roughly insensitive to both $\eta/s$ and initial conditions. Different initial condition models show a big difference of $nMHC(\epsilon_2^2,\epsilon_3^6)$. The ratios totally deviate from 1, which suggests the linear response totally breaks down in this case.
These observables demonstrate that the nonlinear response from final state to initial state will become important for $nMHC(v_2^k,v_3^l)$ with high order of $v_3^l$ ($l>2$). From the results of Fig.~\ref{fig:vnen} (a) and (b) we know that compared with $v_2^k$ to $\epsilon_2^k$, the $v_3^l$ show the faster breakdown of linear correlation to $\epsilon_3^l$ with the increase of centrality or order, and thus $nMHC(v_2^k,v_3^l)$ present such property.
On the other hand, these observables can provide constraints on theoretical models within the framework of hydrodynamics due to their sensitivity to initial conditions or transport coefficients.

With two different initial conditions, {\tt iEBE-VISHNU} models show that $nMHC(v_2^k,v_3^l)<0 $ for $(k+l=4)$, $nMHC(v_2^k,v_3^l)>0 $ for $(k+l=6)$ and $nMHC(v_2^k,v_3^l)<0 $ for $(k+l=8)$ in Fig.~\ref{fig:ordering}.
%More specifically, the ordering behavior among these normalized mixed harmonic multi-particle cumulants can be seen in Fig.~\ref{fig:ordering} (a) and  Fig.~\ref{fig:ordering} (b) that $nMHC(v_2^4,v_3^2)>nMHC(v_2^2,v_3^4)>nMHC(v_2^2,v_3^6)>nMHC(v_2^4,v_3^4)>nMHC(v_2^2,v_3^2)>nMHC(v_2^6,v_3^2)$ . And this behavior is independent of the initial conditions.
The intensities of $nMHC(v_2^k,v_3^l)$ and $nMHC(\epsilon_2^k,\epsilon_3^l)$ with their absolute values are shown in Fig.~\ref{fig:ordering} (c) and Fig.~\ref{fig:ordering} (d) from different initial models respectively. They all show that $\lvert nMHC(v_2^2,v_3^2)\rvert >\lvert nMHC(v_2^2,v_3^4)\rvert >\lvert nMHC(v_2^2,v_3^6)\rvert$ for $(k=2)$ and $\lvert nMHC(v_2^4,v_3^2)\rvert >\lvert nMHC(v_2^4,v_3^4)\rvert$ for $(k=4)$ in the centrality region of 10-50\%, which suggests that with increasing orders of $v_3$, the correlations between $v_2^k$ and $v_3^l$ will become weaker.
The $nMHC(\epsilon_2^k,\epsilon_3^l)$ calculated by initial models also follow these behaviors. It is also testified in Fig.~\ref{fig:nMHC23} that such ordering behaviors are unaffected by the transport coefficients in hydrodynamic evolution, so we hope this property can provide some helps for other collision systems.

In addition to the study of $nMHC$ with higher moments of $v_2$ and $v_3$, we also perform the hydrodynamic calculations of $nMHC$ involving three different flow coefficients, such as $nMHC(v_2^2,v_3^2,v_4^2)$ and $nMHC(v_3^2,v_4^2,v_5^2)$.
In addition to the studies of single $v_n$ fluctuations and the correlations between two flow coefficients, systematic investigations on $nMHC$ with three flow coefficients could bring further insight into the joint probability density function $(p.d.f.)$.
Fig.~\ref{fig:nMHCNL} shows $nMHC(v_m^k,v_n^l)$ and $nMHC(v_m^k,v_n^l,v_p^q)$ calculated by {\tt iEBE-VISHNU} with {\tt TRENTo} and  {\tt AMPT} initial conditions.
We find that $nMHC(v_2^2,v_4^2)$, $nMHC(v_3^2,v_5^2)$, $nMHC(v_4^2,v_5^2)$ and $nMHC(v_2^2, v_3^2,v_4^2)$ are visibly sensitive to both the specific shear viscosity $\eta/s$ and initial conditions. Specifically, they show sizable changes with different $\eta/s$ from {\tt iEBE-VISHNU} with {\tt AMPT} initial condition. Meanwhile, they are influenced by initial conditions, whose corresponding correlators are $nMHC(\epsilon_m^k,\epsilon_n^l)$ or $nMHC(\epsilon_m^k,\epsilon_n^l,\epsilon_p^q)$ for {\tt AMPT} and {\tt TRENTo} initial conditions. While $nMHC(v_3^2, v_4^2)$ calculated by different models give nearly consistent results in non-central collisions.
Besides, for $nMHC(v_2^2,v_4^2)$, $nMHC(v_3^2,v_5^2)$, $nMHC(v_4^2,v_5^2)$ and $nMHC(v_2^2, v_3^2,v_4^2)$ calculated by {\tt iEBE-VISHNU} with {\tt AMPT} initial condition, the results with larger $\eta/s$ tend to give stronger correlations from semi-central to peripheral collisions.
The results of $nMHC(v_2^2,v_4^2)$, $nMHC(v_3^2,v_4^2)$ and $nMHC(v_3^2,v_5^2)$ as well as the $nMHC(v_2^2,v_3^2,v_4^2)$ and $nMHC(v_3^2,v_4^2,v_5^2)$ don't follow the trends of the centrality dependence of the corresponding $nMHC(\epsilon_m^k,\epsilon_n^l)$ and $nMHC(\epsilon_m^k,\epsilon_n^l,\epsilon_p^q)$.
From Fig.~\ref{fig:vnen} (g) we have found that the $v_4$ in the final state is influenced by both $\epsilon_4$ and $\epsilon_2^2$ in the initial conditions, which makes the large deviations between the $nMHC(v_2^2,v_4^2)$ and $nMHC(\epsilon_2^2,\epsilon_4^2)$, especially in the non-central collisions~\cite{Yan:2015jma,Zhu:2016puf}. Similarly, the nonlinear response also affects the other related observables. For example, both $\epsilon_5$ and  $\epsilon_2\epsilon_3$ contribute to the $v_5$ after hydrodynamic evolution, leading to the large deviations of the  magnitudes between $nMHC(v_3^2,v_5^2)$ and $nMHC(\epsilon_3^2,\epsilon_5^2)$. Similar to the case of the $nMHC(v_3^2,v_4^2)$ and $nMHC(\epsilon_3^2,\epsilon_4^2)$, which presents opposite signs in the final states to the initial states in semi-central and peripheral collisions.
Both {\tt AMPT} initial condition and {\tt TRENTo} initial condition give that $nMHC(\epsilon_2^2,\epsilon_3^2,\epsilon_4^2)$ are compatible with zero while the final state correlations $nMHC(v_2^2,v_3^2,v_4^2)$ are negative in non-central collisions from all models. Both {\tt AMPT} initial condition and {\tt TRENTo} initial condition give that $nMHC(\epsilon_3^2,\epsilon_4^2,\epsilon_5^2)$ are non-zero value while the final state correlations $nMHC(v_3^2,v_4^2,v_5^2)$ are consistent with zero from all models. The nonlinear responses of $v_4$ and $v_5$ to $\epsilon_4$ and $\epsilon_5$ play a significantly great role here. If non-zero values of $nMHC(v_3^2,v_4^2,v_5^2)$ are measured in the future experiments, it means that some new mechanisms are involved beyond the current implementation of hydrodynamic model. Future investigations on nonlinear hydrodynamic response with mixed harmonic multi-particle cumulants involving higher flow harmonics will help us to better understand the hydrodynamic evolutions and to provide tighter constrains on the initial models.

\section{Summary}
\label{sec:summary}

We investigate the new proposed observables $nMHC$ in Pb+Pb collisions at $\sqrt{s_{NN}}$=5.02 TeV, by using the event-by-event viscous hydro {\tt iEBE-VISHNU} with {\tt AMPT} and {\tt TRENTo} initial conditions.
The experimental data have not been available yet, so we just present the model predictions here. The calculations were finished before ALICE paper \cite{Acharya:2021hpf} and our results of $nMHC(v_n^k,v_m^l)$ and $MHC(v_n^k,v_m^l,v_p^q)$ were also cited by ALICE paper \cite{Acharya:2021hpf}.
We find $nMHC(v_2^k,v_3^l)$ show little sensitivity to shear viscosity, while $nMHC(v_2^2,v_3^2)$ and $nMHC(v_2^4, v_3^4)$ show visible sensitivity to initial conditions within particular centrality. Besides, the calculations of the ratios $nMHC(v_2^k,v_3^l)$/$nMHC(\epsilon_2^k,\epsilon_3^l)$ demonstrate nonlinear response will become important for $nMHC(v_2^k,v_3^l)$ with high order of $v_3^l$ $(l>2)$. This behavior can be explained by the results of Pearson correlation coefficients namely $r(v_2^k,\epsilon_2^k)$ and $r(v_3^l,\epsilon_3^l)$, which verifies that compared with $v_2^k$ to $\epsilon_2^k$, the $v_3^l$ show the faster breakdown of linear correlation to $\epsilon_3^l$ with the increase of centrality or order.
In addition, we find $nMHC(v_2^k,v_3^l)<0$ for $(k+l=4)$, $nMHC(v_2^k,v_3^l)>0$ for $(k+l=6)$, $nMHC(v_2^k,v_3^l)<0$ for $(k+l=8)$. Considering their intensities by absolute values, we obtain that with increasing orders of $v_3$, the correlations between $v_2^k$ and $v_3^l$ will become weaker. These behaviors are insensitive to the initial condition and shear viscosity, which hopefully provide some helps for other collision systems.
We also calculate the mixed harmonic multi-particle cumulants involving higher-order flow. The $nMHC(v_n^k,v_m^l)$ and $MHC(v_n^k,v_m^l,v_p^q)$ involving higher-order flow show sensitivity to both the specific shear viscosity $\eta/s$ and initial conditions, which can provide further constraints on theoretical models.

\section{Acknowledgments}
M.~L., W.~Z., B.~F, Y.~M. and H.~S. are supported by the NSFC under grant No. 12075007 and No. 11675004.
Y.~Z. is supported by the Villum Young Investigator grant (00025462) from VILLUM FONDEN, the Danish National Research Foundation (Danmarks Grundforskningsfond), and the Carlsberg Foundation (Carlsbergfondet). We acknowledge the computing resources provided by the Super-computing Center of Chinese Academy of Science (SCCAS), Tianhe-1A from the National Super computing Center in Tianjin, China and the High-performance Computing Platform of Peking University.

%
%\clearpage
%-=================================================================
\bibliography{bibliography}
%%-==========================================================================
\end{document}